\documentclass[onecolumn]{aastex61}

\usepackage{amsmath}
\received{November 22, 2017}
\revised{January 29, 2018}
\accepted{January 31, 2018}

\submitjournal{ApJ}

\shorttitle{Gravitational Instability of a Dust Layer} 
\shortauthors{Tatsuuma et al.}

\begin{document}

\title{Gravitational Instability of a Dust Layer Composed of Porous Silicate Dust Aggregates in a Protoplanetary Disk}

\author[0000-0003-1844-5107]{Misako Tatsuuma}
\email{misako.tatsuuma@nao.ac.jp}
\affil{Department of Astronomy, Graduate School of Science, The University of Tokyo, 7-3-1 Hongo, Bunkyo-ku, Tokyo 113-0033, Japan}
\affil{Division of Theoretical Astronomy, National Astronomical Observatory of Japan, 2-21-1 Osawa, Mitaka, Tokyo 181-8588, Japan}

\author{Shugo Michikoshi}
\affil{Department of Contemporary Society, Kyoto Women's University, 35 Kitahiyoshi-cho, Imakumano, Higashiyama-ku, Kyoto 605-8501, Japan}

\author{Eiichiro Kokubo}
\affil{Division of Theoretical Astronomy, National Astronomical Observatory of Japan, 2-21-1 Osawa, Mitaka, Tokyo 181-8588, Japan}
\affil{Department of Astronomy, Graduate School of Science, The University of Tokyo, 7-3-1 Hongo, Bunkyo-ku, Tokyo 113-0033, Japan}

\begin{abstract} 

Planetesimal formation is one of the most important unsolved problems in planet formation theory.
In particular, rocky planetesimal formation is difficult because silicate dust grains are easily broken when they collide.
Recently, it has been proposed that they can grow as porous aggregates when their monomer radius is smaller than $\sim$ 10 nm, which can also avoid the radial drift toward the central star.
However, the stability of a layer composed of such porous silicate dust aggregates has not been investigated.
Therefore, we investigate the gravitational instability of this dust layer.
To evaluate the disk stability, we calculate Toomre's stability parameter $Q$, for which we need to evaluate the equilibrium random velocity of dust aggregates.
We calculate the equilibrium random velocity considering gravitational scattering and collisions between dust aggregates, drag by mean flow of gas, stirring by gas turbulence, and gravitational scattering by gas density fluctuation due to turbulence.
We derive the condition of the gravitational instability using the disk mass, dust-to-gas ratio, turbulent strength, orbital radius, and dust monomer radius.
We find that, for the minimum mass solar nebula model at 1 au, the dust layer becomes gravitationally unstable when the turbulent strength $\alpha\lesssim10^{-5}$.
If the dust-to-gas ratio is increased twice, the gravitational instability occurs for $\alpha\lesssim10^{-4}$.
We also find that the dust layer is more unstable in disks with larger mass, higher dust-to-gas ratio, and weaker turbulent strength, at larger orbital radius, and with a larger monomer radius.

\end{abstract}

\keywords{planets and satellites: formation --- protoplanetary disks --- instabilities --- methods: analytical}

\section{Introduction}\label{sec:intro}

Planetesimal formation is one of the unsolved issues in planet formation theory.
There are several obstacles to the planetesimal formation.
One obstacle is self-induced turbulence \citep[e.g.,][]{Sekiya1998}.
In a protoplanetary disk, sub-$\mathrm{\mu m}$-sized dust grains settle to the disk midplane as they grow.
Such dust settling induces shear instability and then turbulence.
This self-induced turbulence prevents dust grains from settling and a dense dust disk cannot form.
As a result, the gravitational instability (GI) of this dust disk, which leads to rapid planetesimal formation \citep[e.g.,][]{Goldreich1973}, does not occur.
Another obstacle is radial drift \citep[e.g.,][]{Adachi1976}.
In a protoplanetary disk, dust grains orbit around the central star with the Keplerian velocity, while gas orbits with the sub-Keplerian velocity because of the pressure gradient.
Consequently, dust grains experience gas drag, lose their angular momenta, and migrate toward the central star.
They have the fastest drift speed when their Stokes number is unity, corresponding to cm- or m-sized compact bodies with internal density $\sim 1\mathrm{\ g\ cm^{-3}}$.
For 1-m-sized compact bodies at 1 au, the radial drift timescale is around 100 years.
This is much shorter than the disk lifetime $\sim$ several Myr.

Recently, it has been proposed that dust grains become not compact but porous by pairwise accretion \citep[e.g.,][]{Dominik1997,Blum2000,Wada2007,Suyama2008}.
Such porous dust grains, which are called dust aggregates, have fractal dimension $\sim2.5$ \citep{Wada2008} and internal density $\sim10^{-5}$--$10^{-3}\mathrm{\ g\ cm^{-3}}$ \citep{Okuzumi2012}.
They have larger cross-sections than compact dust grains, which means that their collision timescale is shorter.
Also, they have a different law of gas drag from compact dust grains because of their porosity.
In the case of icy dust aggregates with 0.1-$\mathrm{\mu}$m-sized constituent grains, which are called monomers, they can avoid the radial drift and icy planetesimals form \citep{Kataoka2013L}.
During the growth, the GI of the layer of the icy dust aggregates may occur \citep{Michikoshi2016GI}.
However, in contrast to ice, silicate dust grains cannot stick together but, instead, they fragment when they collide \citep[e.g.,][]{Blum1993}.
Silicate dust aggregates with 0.1-$\mathrm{\mu}$m-sized monomers have the critical velocity of catastrophic disruption, which we call the collisional fragmentation velocity, $\sim6\mathrm{\ m\ s^{-1}}$, while icy dust aggregates have $\sim50\mathrm{\ m\ s^{-1}}$ \citep{Wada2009}.

\cite{Arakawa2016} proposed that silicate dust monomers can stick together if they are smaller than $\sim10$ nm because the collisional fragmentation velocity increases with a decreasing monomer radius.
Indeed, it is suggested that dust monomers in a protoplanetary disk are not sub-$\mathrm{\mu}$m-sized interstellar dust grains, but they have experienced evaporation and condensation.
Moreover, some matrix grains in primitive meteorites and interplanetary dust particles contain nm-sized grains \citep{Toriumi1989, Keller2011}.
As nm-sized silicate dust grains grow, they become porous, which can avoid the radial drift.
However, the rocky planetesimal formation mechanism is still unclear, because the stability of the layer consisting of such dust aggregates has not been investigated.
Whether the GI occurs or not is important since the formation process determines the mass and size distributions of planetesimals and thus affects the later formation from planetesimals to planets.

In this paper, we investigate the GI of a dust layer composed of porous dust aggregates of $\sim2.5$--$10$-nm-sized silicate monomers using the method of \cite{Michikoshi2016GI}, which applies the dynamical evolution of planetesimals to such porous dust aggregates.
In Section \ref{sec:model}, we describe models of protoplanetary disks and dust aggregates, and methods to evaluate the stability of the dust layer, which include how to calculate the equilibrium random velocity of dust aggregates.
We present the results in Section \ref{sec:results}.
Finally, Section \ref{sec:sum} is devoted to a summary and discussions.

\section{Models and Methods}\label{sec:model}

To evaluate the stability of the dust layer, we calculate Toomre's stability parameter $Q$ \citep{Toomre1964}, for which we need to evaluate the equilibrium random velocity of dust aggregates.
We describe models of the protoplanetary disk and dust aggregates in Section \ref{subsec:disk} and \ref{subsec:dust}, respectively.
The calculation method of the equilibrium random velocity \citep{Michikoshi2016GI} is presented in Section \ref{subsec:vel}.
Section \ref{subsec:GI} shows conditions of the GI considering the calculated equilibrium random velocity and timescales.

\subsection{Protoplanetary Disks}\label{subsec:disk}

Using the minimum mass solar nebula (MMSN) model \citep{Hayashi1981}, we define the surface densities of gas $\Sigma_\mathrm{g}$ and dust $\Sigma_\mathrm{d}$, and the temperature $T$ as
\begin{eqnarray}
\Sigma_\mathrm{g}&=&1700f_\mathrm{g}\left(\frac{a}{1\ \mathrm{au}}\right)^{-3/2}\mathrm{\ g\ cm^{-2}},\\
\Sigma_\mathrm{d}&=&f_\mathrm{d}\Sigma_\mathrm{g}=1700f_\mathrm{g}f_\mathrm{d}\left(\frac{a}{1\ \mathrm{au}}\right)^{-3/2}\mathrm{\ g\ cm^{-2}},\\
T&=&280\left(\frac{a}{1\ \mathrm{au}}\right)^{-1/2}\mathrm{\ K},
\end{eqnarray}
where $a$ is the orbital radius, $f_\mathrm{g}$ is the ratio to the MMSN model, and $f_\mathrm{d}$ is the dust-to-gas ratio within the H$_2$O snow line.
The MMSN model corresponds to $f_\mathrm{g}=1$ and $f_\mathrm{d}=0.0042$. 
From this temperature profile, we can find that the H$_2$O snow line, where the disk temperature is $T=170$ K, is located at 2.7 au.

We define four disk models as shown in Table \ref{tab:diskmodels} using $f_\mathrm{g}$, $f_\mathrm{d}$, and the dimensionless turbulent strength $\alpha$ \citep[e.g.,][]{Shakura1973}.
Observationally estimated values of $\alpha$ are $10^{-4}\lesssim\alpha\lesssim0.1$ \citep{Andrews2010}, which are derived under the assumption that turbulent viscous diffusion causes the gas disk to accrete onto the central star.
Because smaller $\alpha$ is favorable for the gust growth and the GI, we adopt $10^{-4}$ as the fiducial value.
We assume that the central star is the solar mass, i.e., $M_\ast=M_\odot$.

\begin{table}[htbp]
\centering
\caption{Four disk models} \label{tab:diskmodels}
\begin{tabular}{cccc}
\tablewidth{0pt}
\hline
\hline
Name & $f_\mathrm{g}$ & $f_\mathrm{d}$ & $\alpha$\\
\hline
MMSN & 1 & 0.0042 & $10^{-4}$\\
MMSN weak turbulence & 1 & 0.0042 & $10^{-5}$\\
Massive disk & 2 & 0.0042 & $10^{-4}$\\
Dust-rich disk & 1 & 0.0084 & $10^{-4}$\\
\hline
\end{tabular}
\end{table}

Other disk parameters are as follows.
The isothermal sound speed is
\begin{equation}
c_\mathrm{s}=\sqrt{\frac{k_\mathrm{B}T}{\mu m_\mathrm{H}}}\simeq1.0\times10^5\left(\frac{a}{1\ \mathrm{au}}\right)^{-1/4}\mathrm{\ cm\ s^{-1}},
\end{equation}
where $k_\mathrm{B}$ is the Boltzmann constant, $\mu=2.34$ is the mean molecular weight, and $m_\mathrm{H}$ is the hydrogen mass.
The gas density at the disk midplane is
\begin{equation}
\rho_\mathrm{g}=\frac{\Sigma_\mathrm{g}}{\sqrt{2\pi}c_\mathrm{s}/\Omega_\mathrm{K}}\simeq1.4\times10^{-9}f_\mathrm{g}\left(\frac{a}{1\ \mathrm{au}}\right)^{-11/4}\mathrm{\ g\ cm^{-3}},\label{eq:gasdensity}
\end{equation}
where $\Omega_\mathrm{K}=\sqrt{GM_\ast/a^3}$ is the Keplerian angular velocity and $G$ is the gravitational constant.
The mean free path of gas molecules is
\begin{equation}
l=\frac{\mu m_\mathrm{H}}{\sigma_\mathrm{H_2}\rho_\mathrm{g}}\simeq1.4f_\mathrm{g}^{-1}\left(\frac{a}{1\ \mathrm{au}}\right)^{11/4}\mathrm{\ cm},
\end{equation}
where $\sigma_\mathrm{H_2}=2\times10^{-15}\mathrm{\ cm^2}$ is the collision cross-section of the hydrogen molecule.
The gas-pressure support parameter is
\begin{equation}
\eta=-\frac{1}{2}\left(\frac{c_\mathrm{s}}{a\Omega_\mathrm{K}}\right)^2\frac{\partial\ln(\rho_\mathrm{g}c_\mathrm{s}^2)}{\partial\ln a}\simeq1.8\times10^{-3}\left(\frac{a}{1\ \mathrm{au}}\right)^{1/2}.
\end{equation}
The azimuthal gas velocity is given as $(1-\eta)v_\mathrm{K}$, where $v_\mathrm{K}=a\Omega_\mathrm{K}$ is the Keplerian velocity.
The azimuthal velocity of dust that is decoupled from gas corresponds to $v_\mathrm{K}$.

\subsection{Dust Aggregates}\label{subsec:dust}

We assume that dust aggregates consist of monomers with radius $r_0$ and material density $\rho_0$.
The fiducial radius $r_0=2.5\mathrm{\ nm}$ is selected because \cite{Toriumi1989} found that matrix grains in Allende CV3.2 chondrite have a size distribution with a peak at 5 nm in diameter.
We vary this monomer radius in Section \ref{subsec:resultdust} to investigate the dependence on $r_0$.
The material density of silicate is $\rho_0=3\mathrm{\ g\ cm^{-3}}$.

The static compression pressure $P$ of highly porous dust aggregates is given by \cite{Kataoka2013} as
\begin{equation}
P=\frac{E_\mathrm{roll}}{r_0^3}\left(\frac{\rho_\mathrm{int}}{\rho_0}\right)^3,
\end{equation}
where
\begin{equation}
E_\mathrm{roll}=6\pi^2\gamma r_0\xi=1.1\times10^{-11}\left(\frac{\gamma}{25\mathrm{\ erg\ cm^{-2}}}\right)\left(\frac{r_0}{2.5\mathrm{\ nm}}\right)\left(\frac{\xi}{0.3\mathrm{\ nm}}\right)\mathrm{\ erg}
\end{equation}
is the rolling energy of monomers and $\rho_\mathrm{int}$ is the mean internal density of dust aggregates.
The rolling energy $E_\mathrm{roll}$ is the energy needed to rotate a sphere around an another sphere by $90^\circ$ \citep{Dominik1997}.
To follow \cite{Arakawa2016}, we assume that the surface energy and the critical displacement of silicate are $\gamma=25\mathrm{\ erg\ cm^{-2}}$ and $\xi=0.3$ nm, respectively.
The theoretical critical displacement is $\xi=0.2$ nm \citep{Dominik1997}, while the experimental one is $\xi=3.2$ nm \citep{Heim1999}, and therefore we discuss the uncertainty of $E_\mathrm{roll}$ in Section \ref{sec:sum}.
The self-gravitational pressure $P_\mathrm{grav}$ is given by \cite{Kataoka2013L} as
\begin{equation}
P_\mathrm{grav}=\frac{Gm_\mathrm{d}^2/r_\mathrm{d}^2}{\pi r_\mathrm{d}^2}=\frac{Gm_\mathrm{d}^2}{\pi r_\mathrm{d}^4},
\end{equation}
where $m_\mathrm{d}$ and $r_\mathrm{d}$ are mass and radius of dust aggregates, respectively.

We assume that a dust aggregate has a spherical body, and thus the relationship among $m_\mathrm{d}$, $r_\mathrm{d}$, and $\rho_\mathrm{int}$ is given as $m_\mathrm{d}=(4/3)\pi r_\mathrm{d}^3\rho_\mathrm{int}$.
For simplicity, we do not consider the size distribution of monomers $r_0$ and assume that all dust aggregates have the identical mass $m_\mathrm{d}$ with the mean internal density $\rho_\mathrm{int}$. 
We derive the equation of dust evolution via quasi-static self-gravitational compression by equating $P$ and $P_\mathrm{grav}$, which is described as
\begin{equation}
\rho_\mathrm{int}=\left(\frac{r_0^3}{E_\mathrm{roll}}\frac{Gm_\mathrm{d}^2}{\pi r_\mathrm{d}^4}\right)^{1/3}\rho_0
=0.14\left(\frac{G}{\gamma\xi}\right)^{3/5}r_0^{6/5}\rho_0^{9/5}m_\mathrm{d}^{2/5} \label{eq:evoltrack}.
\end{equation}
This self-gravitational compression dominated other compression mechanisms when $m_\mathrm{d}\gtrsim10^{13}$ g \citep{Arakawa2016}.

\subsection{Random Velocity}\label{subsec:vel}

To calculate the equilibrium random velocity of dust aggregates $v$, we divide $\mathrm{d}v^2/\mathrm{d}t$ into five components, which is given as
\begin{equation}
\frac{\mathrm{d}v^2}{\mathrm{d}t}=\left(\frac{\mathrm{d}v^2}{\mathrm{d}t}\right)_\mathrm{grav}+\left(\frac{\mathrm{d}v^2}{\mathrm{d}t}\right)_\mathrm{col}+\left(\frac{\mathrm{d}v^2}{\mathrm{d}t}\right)_\mathrm{gas,drag}+\left(\frac{\mathrm{d}v^2}{\mathrm{d}t}\right)_\mathrm{turb,stir}+\left(\frac{\mathrm{d}v^2}{\mathrm{d}t}\right)_\mathrm{turb,scat}=0.
\label{eq:randomvel}
\end{equation}
Each component represents from left to right gravitational scattering between dust aggregates, collisions between them, drag by mean flow of gas, stirring by gas turbulence, and gravitational scattering by gas density fluctuation due to turbulence.
We assume that the velocity distribution is isotropic, i.e., $v_x\simeq v_y\simeq v_z\simeq v/\sqrt{3}$, where $v_x$, $v_y$, and $v_z$ are $x$, $y$, and $z$ components of $v$, respectively.
In reality, the velocity distribution is anisotropic, but the effects are not significant \citep{Michikoshi2017}.

\subsubsection{Dust-Dust Interaction}

The velocity change by gravitational scattering between dust aggregates is given as
\begin{equation}
\left(\frac{\mathrm{d}v^2}{\mathrm{d}t}\right)_\mathrm{grav}=n_\mathrm{d}\pi\left(\frac{2Gm_\mathrm{d}}{v^2_\mathrm{rel}}\right)^2v_\mathrm{rel}v^2\ln\Lambda,
\label{eq:grav}
\end{equation}
which is derived from the Chandrasekhar's two-body relaxation time \citep[e.g.,][]{Ida1990}.
The number density of dust aggregates is
\begin{equation}
n_\mathrm{d}\simeq\frac{\Sigma_\mathrm{d}/m_\mathrm{d}}{\sqrt{2\pi}v_z/\Omega_\mathrm{K}},
\end{equation}
the typical relative velocity between them is $v_\mathrm{rel}\simeq\sqrt{2}v$, and $\Lambda$ is defined as
\begin{equation}
\Lambda=v_\mathrm{rel}^2\frac{v_z/\Omega_\mathrm{K}+r_\mathrm{H}}{2Gm_\mathrm{d}},
\end{equation}
where
\begin{equation}
r_\mathrm{H}=\left(\frac{2m_\mathrm{d}}{3M_\ast}\right)^{1/3}a
\end{equation}
is the Hill radius \citep{Stewart2000}.
In equation (\ref{eq:grav}), $\pi(2Gm_\mathrm{d}/v_\mathrm{rel}^2)^2$ means the gravitational scattering cross-section.

The velocity change by collisions between dust aggregates is given as
\begin{equation}
\left(\frac{\mathrm{d}v^2}{\mathrm{d}t}\right)_\mathrm{col}=-C_\mathrm{col}n_\mathrm{d}\pi(2r_\mathrm{d})^2\left(1+\frac{v_\mathrm{esc}^2}{v_\mathrm{rel}^2}\right)v_\mathrm{rel}v^2,
\label{eq:col}
\end{equation}
where $C_\mathrm{col}=1/2$ \citep[e.g.,][]{Inaba2001} is the rate of change of kinetic energy during a collision and $v_\mathrm{esc}=\sqrt{2Gm_\mathrm{d}/r_\mathrm{d}}$ is the surface escape velocity.
In equation (\ref{eq:col}), $\pi(2r_\mathrm{d})^2(1+v_\mathrm{esc}^2/v_\mathrm{rel}^2)$ means the collision cross-section including gravitational focusing.
We assume that results of all collisions are accretion.

\subsubsection{Dust-Gas Interaction}

Drag by mean flow of gas is given as
\begin{equation}
\left(\frac{\mathrm{d}v^2}{\mathrm{d}t}\right)_\mathrm{gas,drag}=-\frac{2}{t_\mathrm{s}}v^2,
\label{eq:gasdrag}
\end{equation}
where 
\begin{equation}
t_\mathrm{s}=\frac{2m_\mathrm{d}}{\pi C_\mathrm{D}r_\mathrm{d}^2\rho_\mathrm{g}u}
\label{eq:stoptime}
\end{equation}
is the stopping time, $C_\mathrm{D}$ is the dimensionless drag coefficient, and $u\simeq\sqrt{v^2+\eta^2v_\mathrm{K}^2}$ is the relative velocity between dust and gas \citep[e.g.,][]{Adachi1976}.
In the case of $m_\mathrm{g}\gtrsim10^{13}$, we confirm that the Stokes number $\tau_\mathrm{s}=\Omega_\mathrm{K} t_\mathrm{s}\gg1$ (see Figure \ref{fig:St}), and therefore the dust aggregates are decoupled from gas.
We adopt the expression of $C_\mathrm{D}$ from \cite{Brown2003}
\begin{equation}
C_\mathrm{D}=\begin{cases}
\cfrac{8v_\mathrm{th}}{3u} & (r_\mathrm{d}<9l/4)\\
\cfrac{0.407}{1+8710/\mathrm{Re}}+\cfrac{24}{\mathrm{Re}}(1+0.150\mathrm{Re}^{0.681}) & (r_\mathrm{d}>9l/4)
\end{cases},
\label{eq:CD}
\end{equation}
where $v_\mathrm{th}=\sqrt{8/\pi}c_\mathrm{s}$ is the thermal velocity, $\mathrm{Re}=2r_\mathrm{d}u/\nu$ is the Reynolds number, and $\nu=v_\mathrm{th}l/2$ is the viscosity.
When dust aggregates are smaller than the mean free path of gas molecules $(r_\mathrm{d}<9l/4)$, drag felt by dust aggregates is the Epstein drag.
In the other regime $(r_\mathrm{d}>9l/4)$, there are the Stokes and the Newton drag for low and high Reynolds numbers, respectively.

Stirring by gas turbulence is given as
\begin{equation}
\left(\frac{\mathrm{d}v^2}{\mathrm{d}t}\right)_\mathrm{turb,stir}=\frac{2\tau_\mathrm{e}v_\mathrm{t}^2\Omega_\mathrm{K}}{\tau_\mathrm{s}(\tau_\mathrm{e}+\tau_\mathrm{s})},
\label{eq:turbstir}
\end{equation}
which is derived from the equilibrium velocity by turbulent stirring $v_\mathrm{t}^2\tau_\mathrm{e}/(\tau_\mathrm{e}+\tau_\mathrm{s})$ \citep{Youdin2007}.
The dimensionless eddy turnover time is $\tau_\mathrm{e}=1$ \citep[e.g.,][]{Youdin2011} and the turbulent velocity is $v_\mathrm{t}=\sqrt{\alpha}c_\mathrm{s}$.

Gravitational scattering by gas density fluctuation due to turbulence changes the velocity as
\begin{equation}
\left(\frac{\mathrm{d}v^2}{\mathrm{d}t}\right)_\mathrm{turb,scat}=C_\mathrm{turb}\alpha\left(\frac{\Sigma_\mathrm{g}a^2}{M_\ast}\right)^2\Omega_\mathrm{K}^3a^2,
\label{eq:turbscat}
\end{equation}
which is derived by \cite{Okuzumi2013}.
The dimensionless coefficient determined by disk structure $C_\mathrm{turb}$ is given as
\begin{equation}
C_\mathrm{turb} = \frac{0.94{\cal{L}}}{(1+4.5H_\mathrm{res,0}/H)^2},
\end{equation}
where $\cal{L}$ is the saturation limiter, $H_\mathrm{res,0}$ is the vertical dead zone half width, and $H$ is the gas scale height.
We adopt ${\cal{L}}=1$, which means that turbulence occurs due to the magneto-rotational instability (MRI).
We use $C_\mathrm{turb}=3.1\times10^{-2}$ by assuming $H_\mathrm{res,0}=H$.

\subsection{GI Conditions}\label{subsec:GI}

\subsubsection{Toomre's $Q$}

We define the condition of the GI using Toomre's stability parameter $Q$ \citep{Toomre1964}
\begin{equation}
Q = \frac{v_x\Omega_\mathrm{K}}{3.36G\Sigma_\mathrm{d}},
\label{eq:Q}
\end{equation}
where $v_x$ is derived from equation (\ref{eq:randomvel}).
The axisymmetric mode grows when $Q<1$ \citep{Toomre1964}.
When $1\lesssim Q\lesssim2$, non-axisymmetric mode or self-gravity wakes grow \citep[e.g.,][]{Toomre1981}, and planetesimals are formed in these self-gravity wakes \citep[e.g.,][]{Michikoshi2007}.
Therefore, we define the condition of the GI as $Q<2$.

\subsubsection{Timescales}

We compare timescales of growth by pairwise accretion $t_\mathrm{grow}$, the radial drift $t_\mathrm{drift}$, and the GI $t_\mathrm{GI}$.
The substantial radial drift occurs when $t_\mathrm{grow}>(1/30)t_\mathrm{drift}$ \citep{Okuzumi2012}.
The growth timescale considering gravitational focusing is given as 
\begin{equation}
t_\mathrm{grow}\equiv\frac{m_\mathrm{d}}{\mathrm{d}m_\mathrm{d}/\mathrm{d}t}=\frac{1}{n_\mathrm{d}\pi r_\mathrm{d}^2(1+v_\mathrm{esc}^2/v_\mathrm{rel}^2)v_\mathrm{rel}},\label{eq:tgrow}
\end{equation}
and the radial drift timescale due to the gas drag \citep[e.g.,][]{Adachi1976} is given as
\begin{equation}
t_\mathrm{drift}\equiv\frac{a}{\mathrm{d}a/\mathrm{d}t}=\frac{a}{2\tau_\mathrm{s}\eta v_\mathrm{K}/(1+\tau_\mathrm{s}^2)}.\label{eq:tdrift}
\end{equation}
The GI timescale \citep[e.g.,][]{Sekiya1983,Goldreich1973} is on the order of orbital period
\begin{equation}
t_\mathrm{GI}\sim\Omega_\mathrm{K}^{-1}.
\label{eq:tGI}
\end{equation}
If the GI timescale is the shortest when $Q<2$, we conclude that the GI occurs.

\subsubsection{Velocity Range}

In order for the GI to occur in our models, we confirm that runaway growth and fragmentation do not occur when $Q<2$.
The dust aggregates cannot grow if the fragmentation occurs, and the assumption of no mass distribution is broken if the runaway growth occurs.
Conditions of the runaway growth and the fragmentation are $v_\mathrm{rel}<v_\mathrm{esc}$ \citep{Kokubo1996} and $v_\mathrm{rel}>v_\mathrm{frag,cr}$, respectively, where
\begin{equation}
v_\mathrm{frag,cr}=6\times10^2\left(\frac{r_0}{100\mathrm{\ nm}}\right)^{-5/6}\mathrm{\ cm\ s^{-1}}
\label{eq:frag}
\end{equation}
is the critical velocity of catastrophic disruption \citep{Dominik1997,Wada2009}.
Therefore, $v_\mathrm{esc}<v_\mathrm{rel}<v_\mathrm{frag,cr}$ is needed if dust aggregates are not to experience the runaway growth and the fragmentation.

\section{Results}\label{sec:results}

First, we investigate the stability of the dust layer for the fiducial model, whose orbital radius is 1 au and dust monomer radius is $r_0=2.5$ nm in Section \ref{subsec:resultQ}, \ref{subsec:resulttime}, and \ref{subsec:resultvel}.
We use the four disk models shown in Table \ref{tab:diskmodels}.
Next, we calculate the dependence on parameters in Section \ref{subsec:resultdep}.

\subsection{Toomre's $Q$}\label{subsec:resultQ}

\begin{figure}[htbp]
\plotone{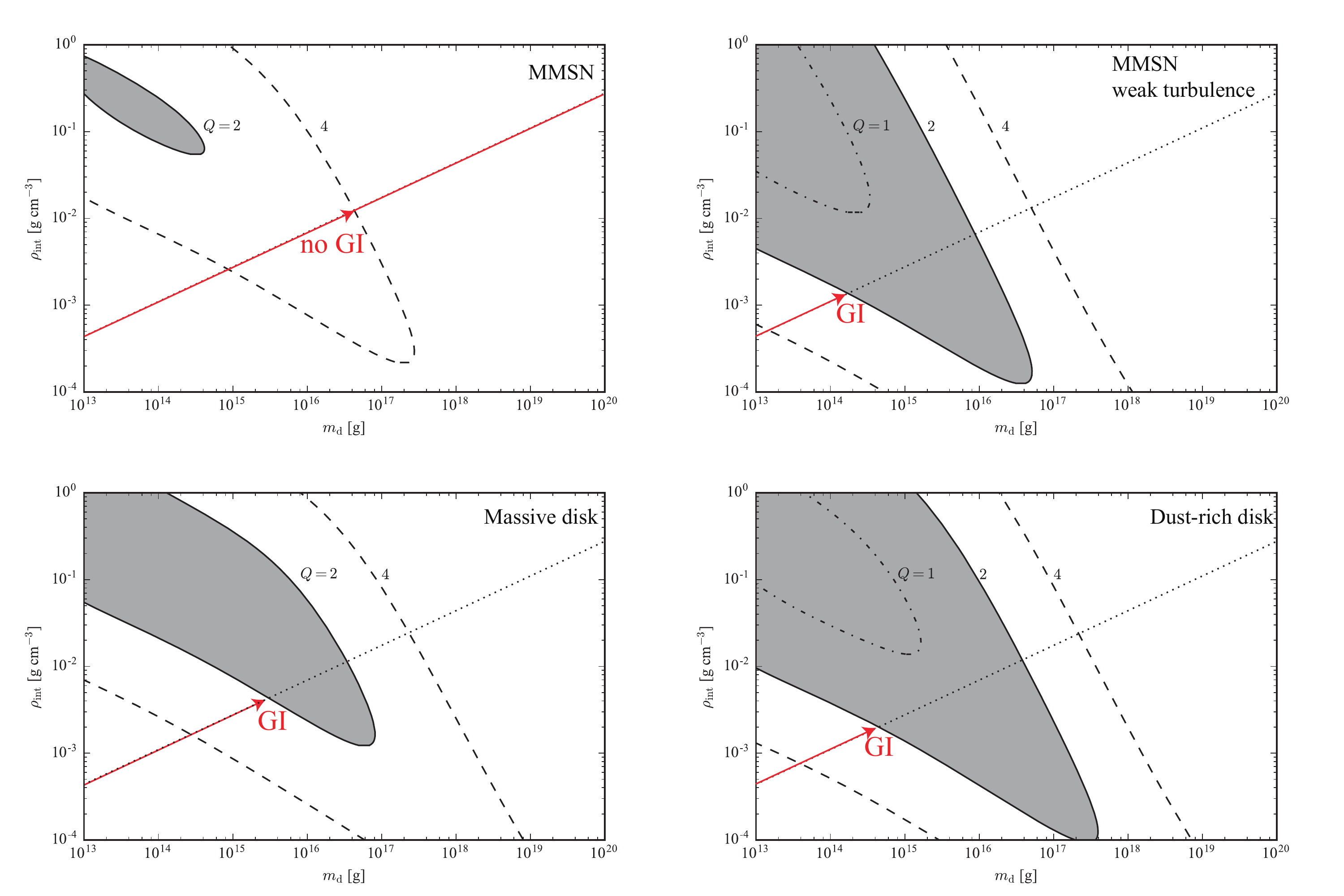}
\caption{Toomre's $Q$ in the $m_\mathrm{d}$-$\rho_\mathrm{int}$ plane at 1 au of the MMSN (top left), MMSN weak turbulence (top right), massive (bottom left), and dust-rich disk (bottom right) models.
The dash-dotted, solid, and dash contours correspond to $Q=1$, 2, and 4, respectively.
The dotted lines and the red arrows show the mass and internal density relation under self-gravitational compression of dust aggregates with $r_0=2.5$ nm.}
\label{fig:ToomresQ}
\end{figure}

We calculate Toomre's $Q$ at 1 au of four disk models shown in Table \ref{tab:diskmodels} and draw contours in the $m_\mathrm{d}$-$\rho_\mathrm{int}$ plane in Figure \ref{fig:ToomresQ}.
Also, we show the mass and internal density relation under the self-gravitational compression of dust aggregates with $r_0=2.5$ nm using equation (\ref{eq:evoltrack}).
All models show a tendency for $Q$ to decrease and then increase as dust aggregates grow.
The GI occurs in three of them: the MMSN weak turbulence, massive, and dust-rich disk models.

\begin{figure}[htbp]
\plotone{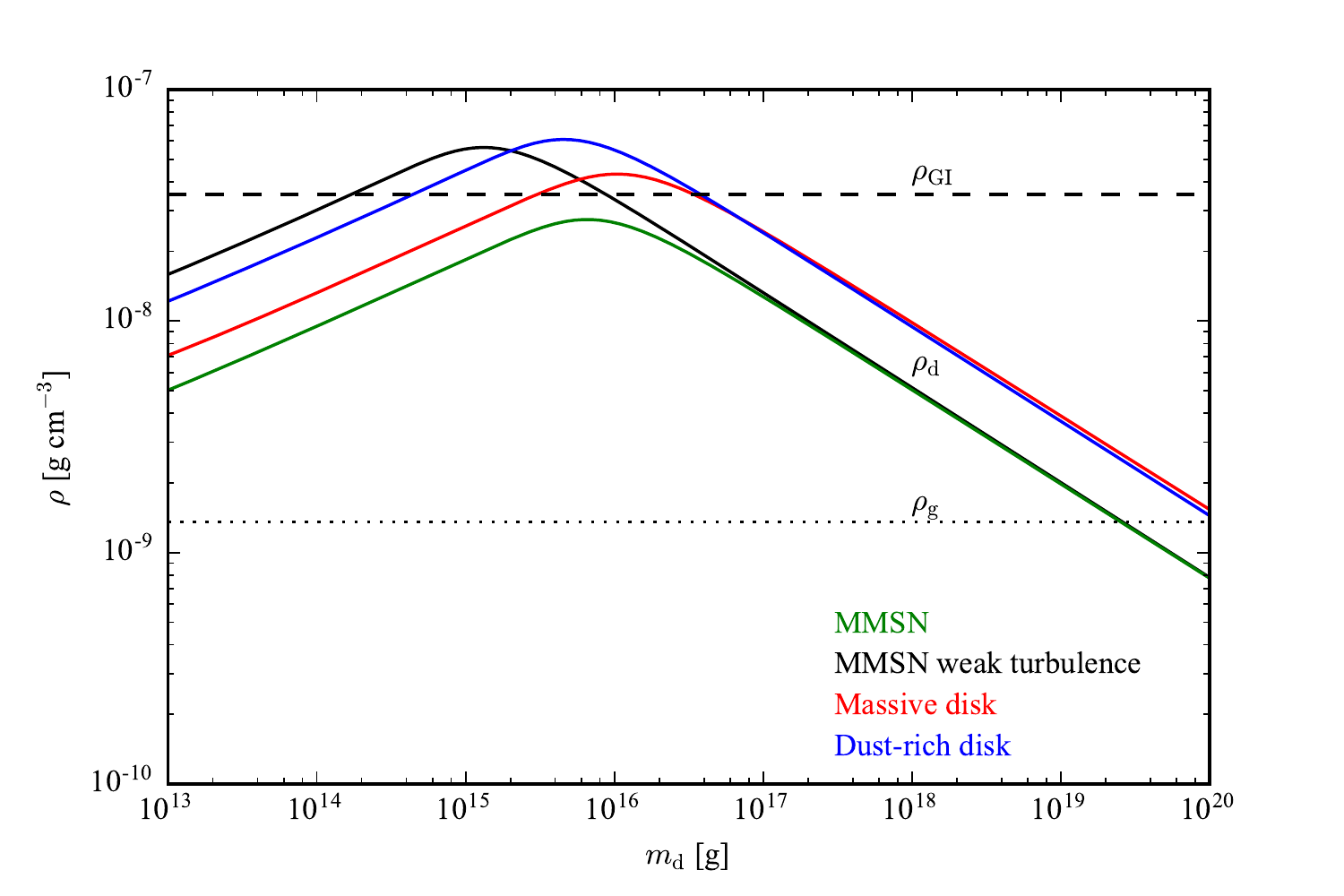}
\caption{Densities of dust aggregates $\rho_\mathrm{d}$ (solid) and gas $\rho_\mathrm{g}$ (dot), and the critical density for the GI $\rho_\mathrm{GI}$ (dash) against $m_\mathrm{d}$ at 1 au of the MMSN (green), MMSN weak turbulence (black), massive (red), and dust-rich (blue) disk models of dust aggregates with $r_0=2.5$ nm.}
\label{fig:massdensity}
\end{figure}

We also plot the densities of dust aggregates and gas at 1 au for four disk models in Figure \ref{fig:massdensity}, where the density of dust aggregates is given as
\begin{equation}
\rho_\mathrm{d} \simeq n_\mathrm{d}m_\mathrm{d}.
\end{equation}
Note that the dust internal density $\rho_\mathrm{int}$ is determined by self-gravitational compression.
The critical density for the GI, which is calculated from equation (\ref{eq:Q}) as
\begin{equation}
\rho_\mathrm{GI}\simeq\frac{\Omega_\mathrm{K}^2}{3.36\sqrt{2\pi}QG}\simeq3.5\times10^{-8}\left(\frac{Q}{2}\right)^{-1}\left(\frac{a}{\mathrm{1\ au}}\right)^{-3}\mathrm{\ g\ cm^{-3}},
\end{equation}
is also shown in Figure \ref{fig:massdensity}.
It is found that the dust-to-gas ratio at the disk midplane is $\sim26$ when the GI occurs.

\subsection{Timescales}\label{subsec:resulttime}

\begin{figure}[htbp]
\plotone{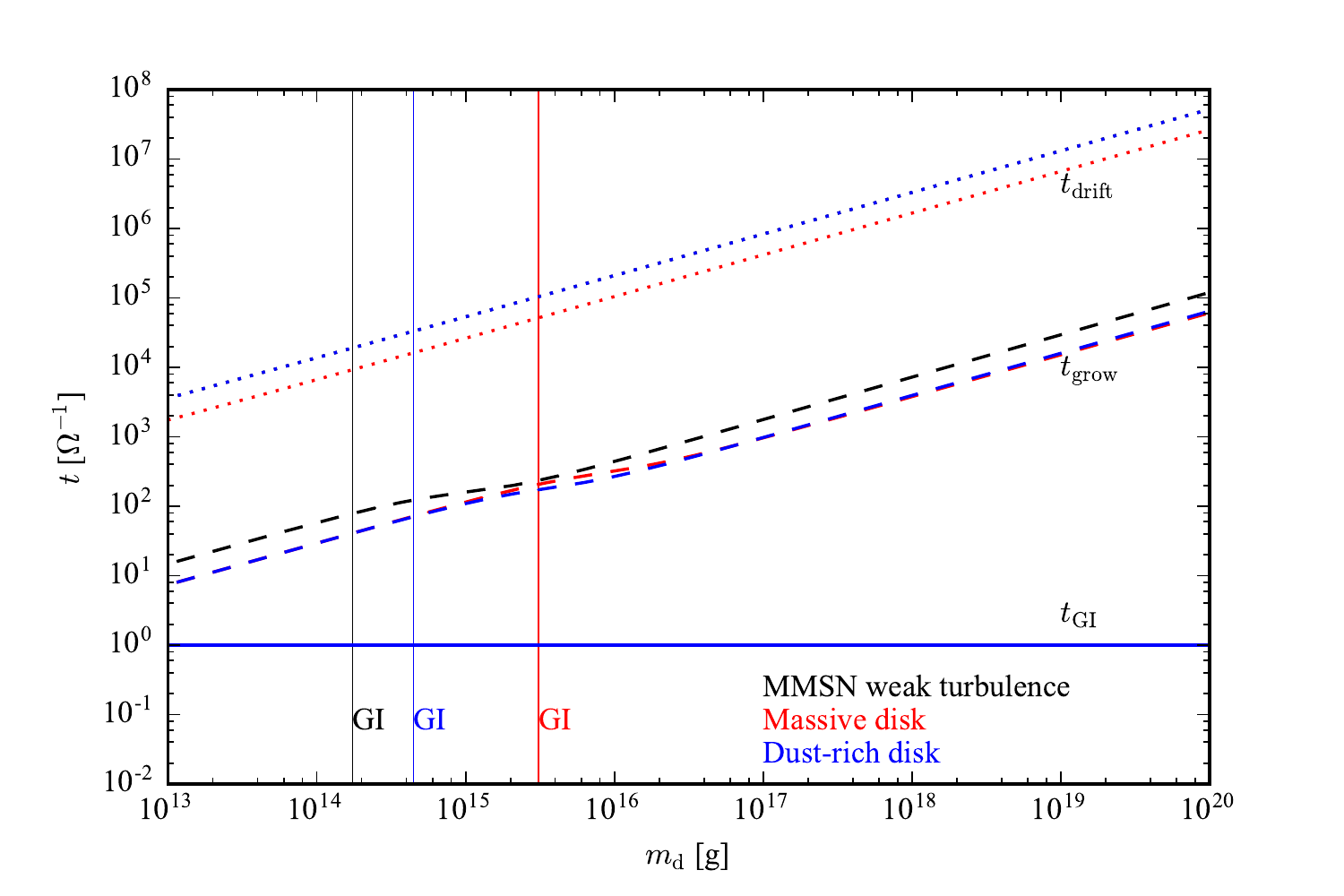}
\caption{Timescales of the growth $t_\mathrm{grow}$ (dash), the radial drift $t_\mathrm{drift}$ (dot), and the GI $t_\mathrm{GI}$ (solid) against $m_\mathrm{d}$ at 1 au of the MMSN weak turbulence (black), massive (red), and dust-rich (blue) disk models of dust aggregates with $r_0=2.5$ nm.
The vertical lines show when the GI occurs.}
\label{fig:timescale}
\end{figure}

Timescales of the growth $t_\mathrm{grow}$ (equation (\ref{eq:tgrow})), the radial drift $t_\mathrm{drift}$ (equation (\ref{eq:tdrift})), and the GI $t_\mathrm{GI}$ (equation (\ref{eq:tGI})) at 1 au of the three models where the GI occurs are shown in Figure \ref{fig:timescale}.
We use equation (\ref{eq:evoltrack}) with $r_0=2.5$ nm.
The growth timescale $t_\mathrm{grow}\propto m_\mathrm{d}^{3/5}$ and the radial drift timescale $t_\mathrm{drift}\propto m_\mathrm{d}^{3/5}$ increase with $m_\mathrm{d}$ monotonically, while the GI timescale $t_\mathrm{GI}$ is independent of $m_\mathrm{d}$.
At $Q=2$, the shortest timescale of all models is the GI.

\subsection{Equilibrium Random Velocity}\label{subsec:resultvel}

\begin{figure}[htbp]
\plotone{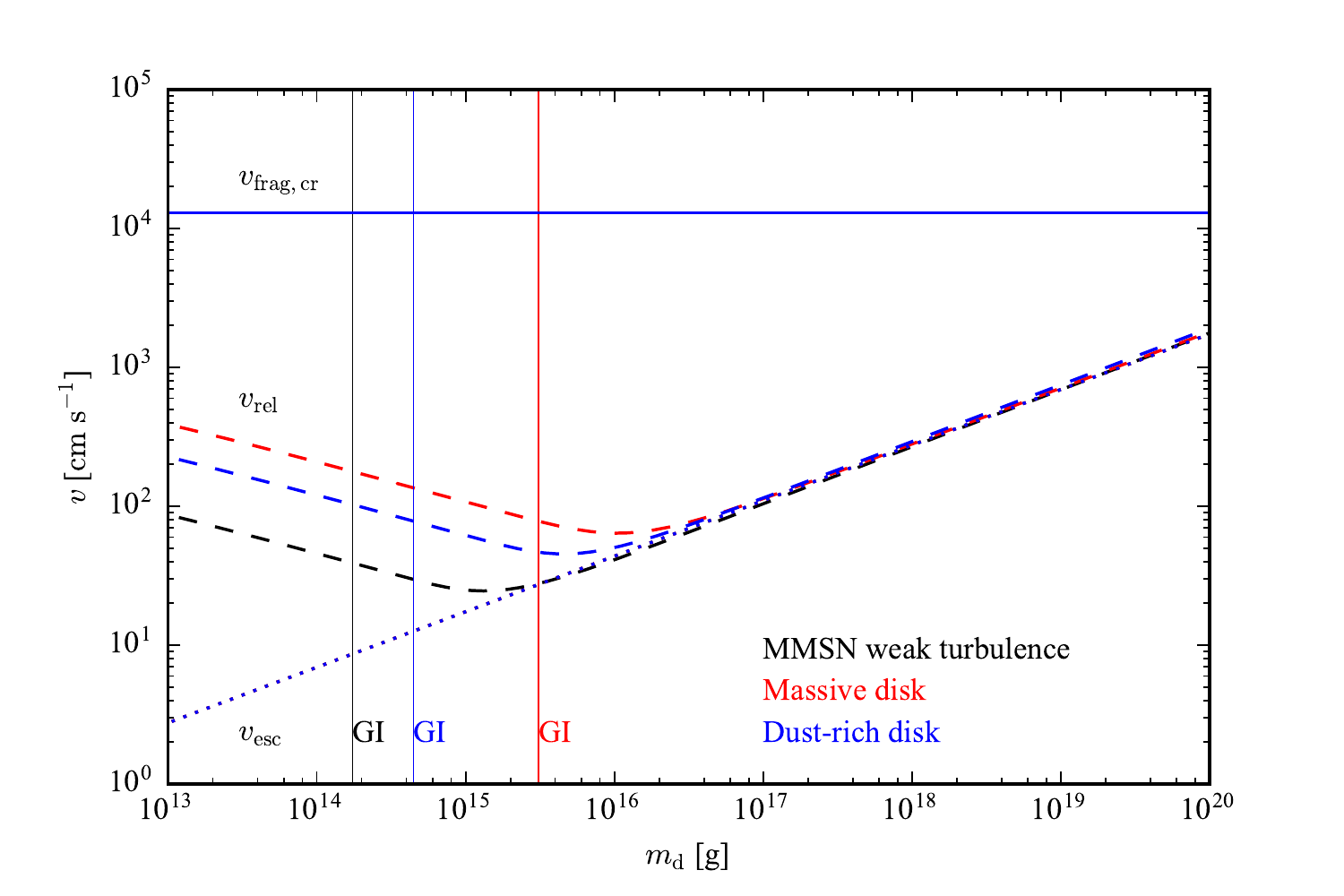}
\caption{Relative velocity between dust aggregates $v_\mathrm{rel}$ (dash), the escape velocity $v_\mathrm{esc}$ (dot), and the fragmentation velocity $v_\mathrm{frag,cr}$ (solid) against $m_\mathrm{d}$ at 1 au of the MMSN weak turbulence (black), massive (red), and dust-rich (blue) disk models of dust aggregates with $r_0=2.5$ nm.
The vertical lines show when the GI occurs.}
\label{fig:velocity}
\end{figure}

To check that the runaway growth and the fragmentation do not occur, we plot $v_\mathrm{rel}\simeq\sqrt{2}v$, $v_\mathrm{esc}=\sqrt{2Gm_\mathrm{d}/r_\mathrm{d}}\propto m_\mathrm{d}^{2/5}$, and $v_\mathrm{frag,cr}$ (equation (\ref{eq:frag})) at 1 au of the three models where the GI occurs in Figure \ref{fig:velocity}.
Obviously, $v_\mathrm{esc}$ and $v_\mathrm{frag,cr}$ are independent of disk models.
For small $m_\mathrm{d}$, the condition of $v_\mathrm{frag,cr}>v_\mathrm{rel}>v_\mathrm{esc}$ is satisfied, which means that the runaway growth and the fragmentation do not occur.
When the GI occurs, $v_\mathrm{rel}$ is still larger than $v_\mathrm{esc}$.

\begin{figure}[htbp]
\plotone{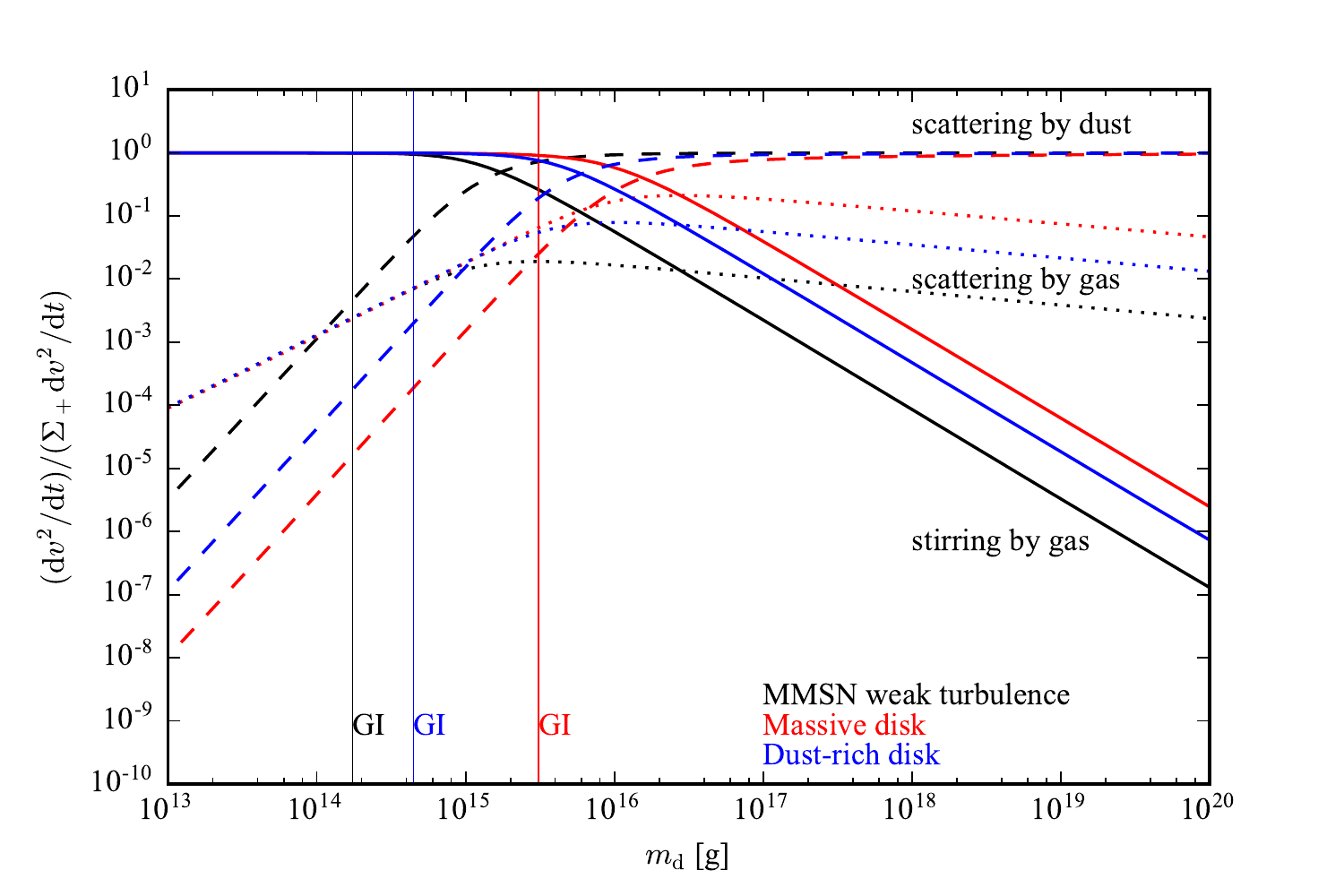}
\caption{The relative contribution of each increasing mechanism in $\mathrm{d}v^2/\mathrm{d}t$, which includes gravitational scattering between dust aggregates (dash), stirring by gas turbulence (solid), and gravitational scattering by gas density fluctuation due to turbulence (dot) at 1 au of the MMSN weak turbulence (black), massive (red), and dust-rich (blue) disk models of dust aggregates with $r_0=2.5$ nm.
Each $\mathrm{d}v^2/\mathrm{d}t$ is divided by $\Sigma_+\mathrm{d}v^2/\mathrm{d}t=(\mathrm{d}v^2/\mathrm{d}t)_{\mathrm{grav}}+(\mathrm{d}v^2/\mathrm{d}t)_{\mathrm{turb,stir}}+(\mathrm{d}v^2/\mathrm{d}t)_{\mathrm{turb,scat}}$.
The vertical lines show when the GI occurs.}
\label{fig:heat}
\end{figure}

Figure \ref{fig:heat} shows the relative contribution of each increasing mechanism in $\mathrm{d}v^2/\mathrm{d}t$ of each increasing mechanism, which includes gravitational scattering between dust aggregates (equation (\ref{eq:grav})), stirring by gas turbulence (equation (\ref{eq:turbstir})), and gravitational scattering by gas density fluctuation due to turbulence (equation (\ref{eq:turbscat})).
The main increasing mechanism is stirring by gas turbulence before the GI occurs.
Gravitational scattering dominates others when $m_\mathrm{d}\gtrsim10^{15}$ g.

\begin{figure}[htbp]
\plotone{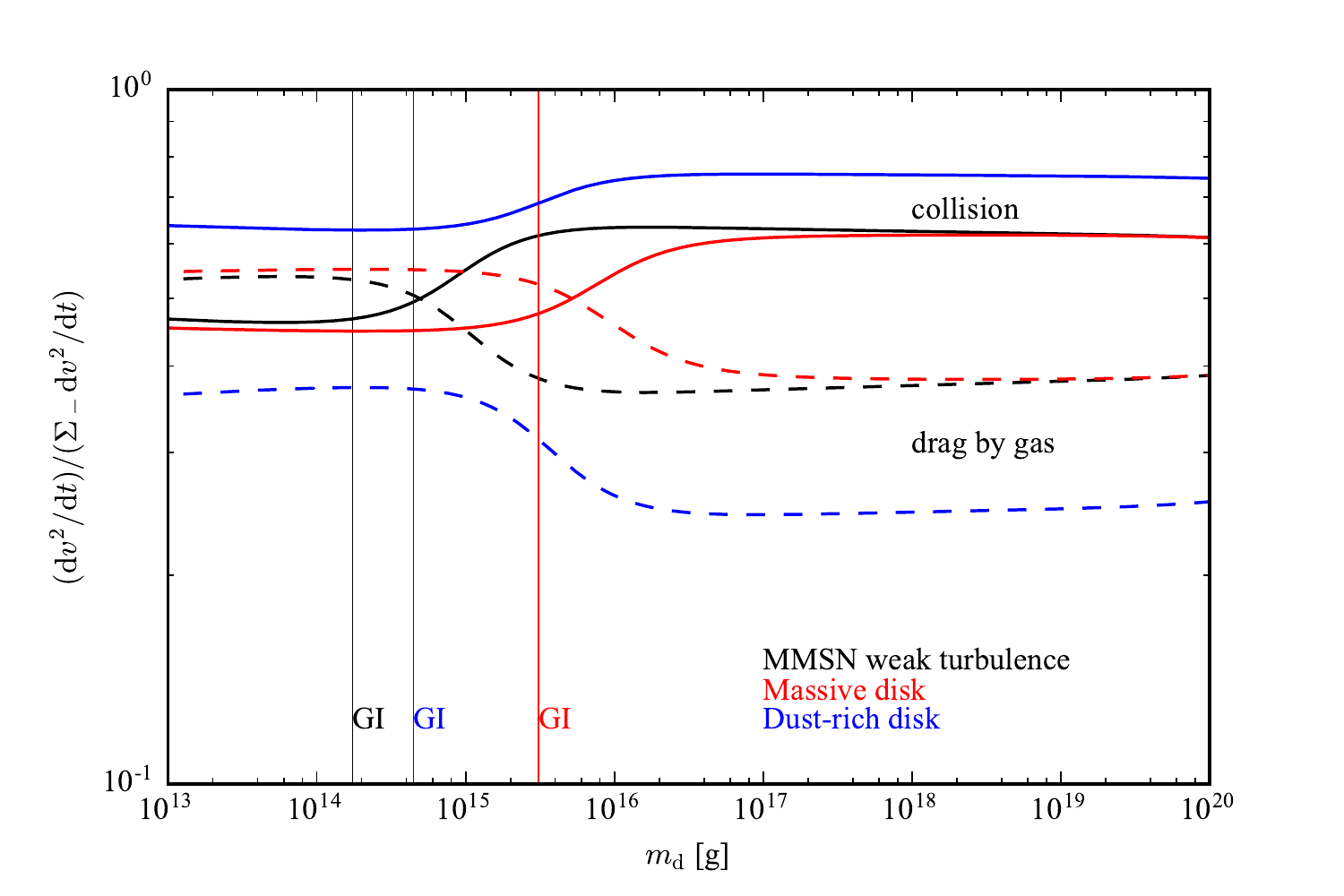}
\caption{The relative contribution of each decreasing mechanism in $\mathrm{d}v^2/\mathrm{d}t$, which includes collisions between dust aggregates (solid) and drag by mean flow of gas (dash) at 1 au of the MMSN weak turbulence (black), massive (red), and dust-rich (blue) disk models of dust aggregates with $r_0=2.5$ nm.
Each $\mathrm{d}v^2/\mathrm{d}t$ is divided by $\Sigma_-\mathrm{d}v^2/\mathrm{d}t=(\mathrm{d}v^2/\mathrm{d}t)_{\mathrm{col}}+(\mathrm{d}v^2/\mathrm{d}t)_{\mathrm{gas,drag}}$.
The vertical lines show when the GI occurs.}
\label{fig:cool}
\end{figure}

Figure \ref{fig:cool} is the same as Figure \ref{fig:heat} but for the decreasing mechanism, which includes collisions between dust aggregates (equation (\ref{eq:col})) and drag by mean flow of gas (equation (\ref{eq:gasdrag})).
The difference between the two decreasing mechanisms is a few factors, which means that they have comparable effects.

\begin{figure}[htbp]
\plotone{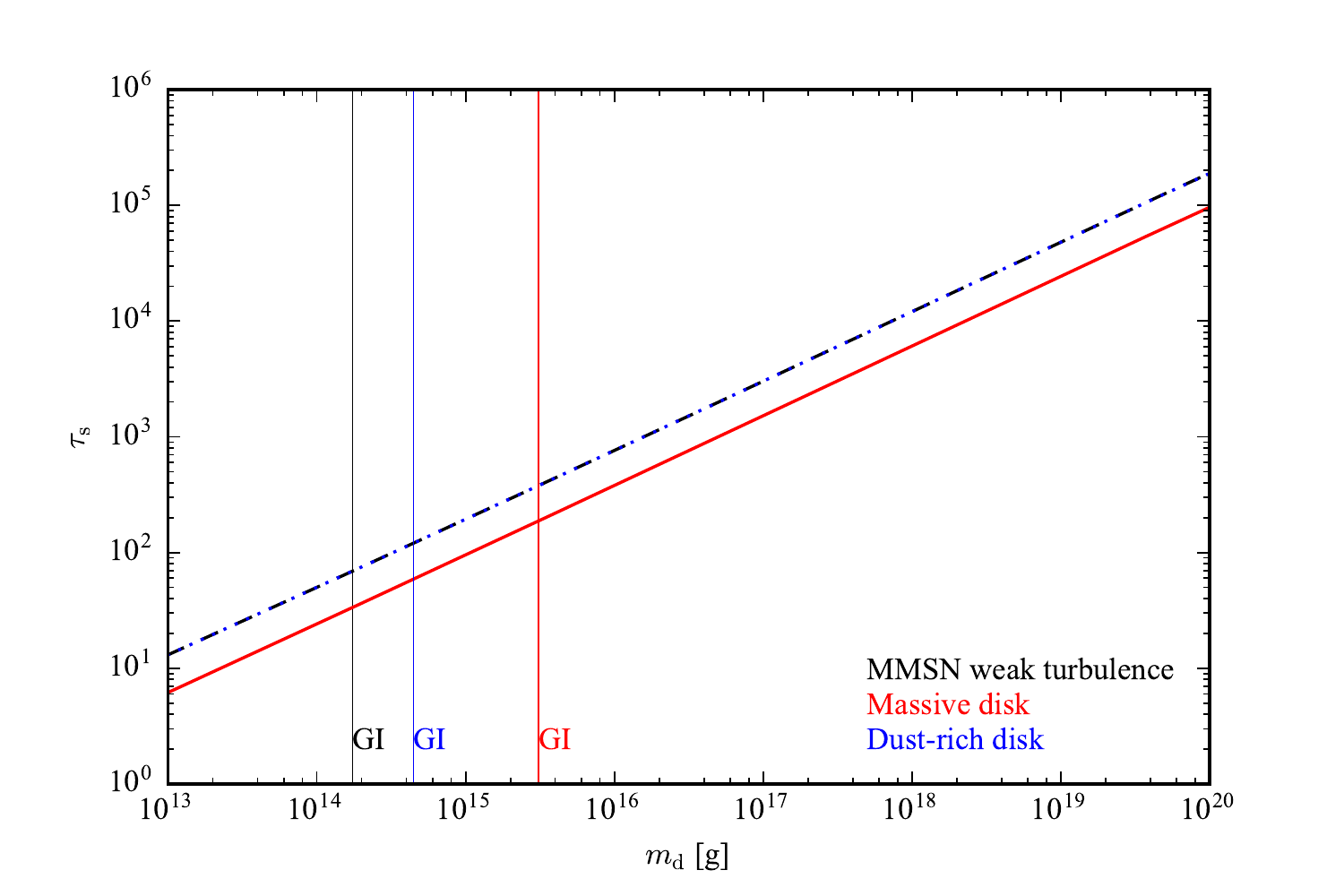}
\caption{The Stokes numbers $\tau_\mathrm{s}$ at 1 au of the MMSN weak turbulence (black-dashed), massive (red-solid), and dust-rich (blue-dotted) disk models of dust aggregates with $r_0=2.5$ nm.
The vertical lines show when the GI occurs.}
\label{fig:St}
\end{figure}

In addition, we plot the Stokes number $\tau_\mathrm{s}$ to show the effect of coupling between dust and gas in Figure \ref{fig:St}.
The MMSN weak turbulence and dust-rich disk models are not distinguishable.
The Stokes number is always much larger than unity. 

\subsection{Dependence on Parameters}\label{subsec:resultdep}

\subsubsection{Disk Parameters}

\begin{figure}[htbp]
\plotone{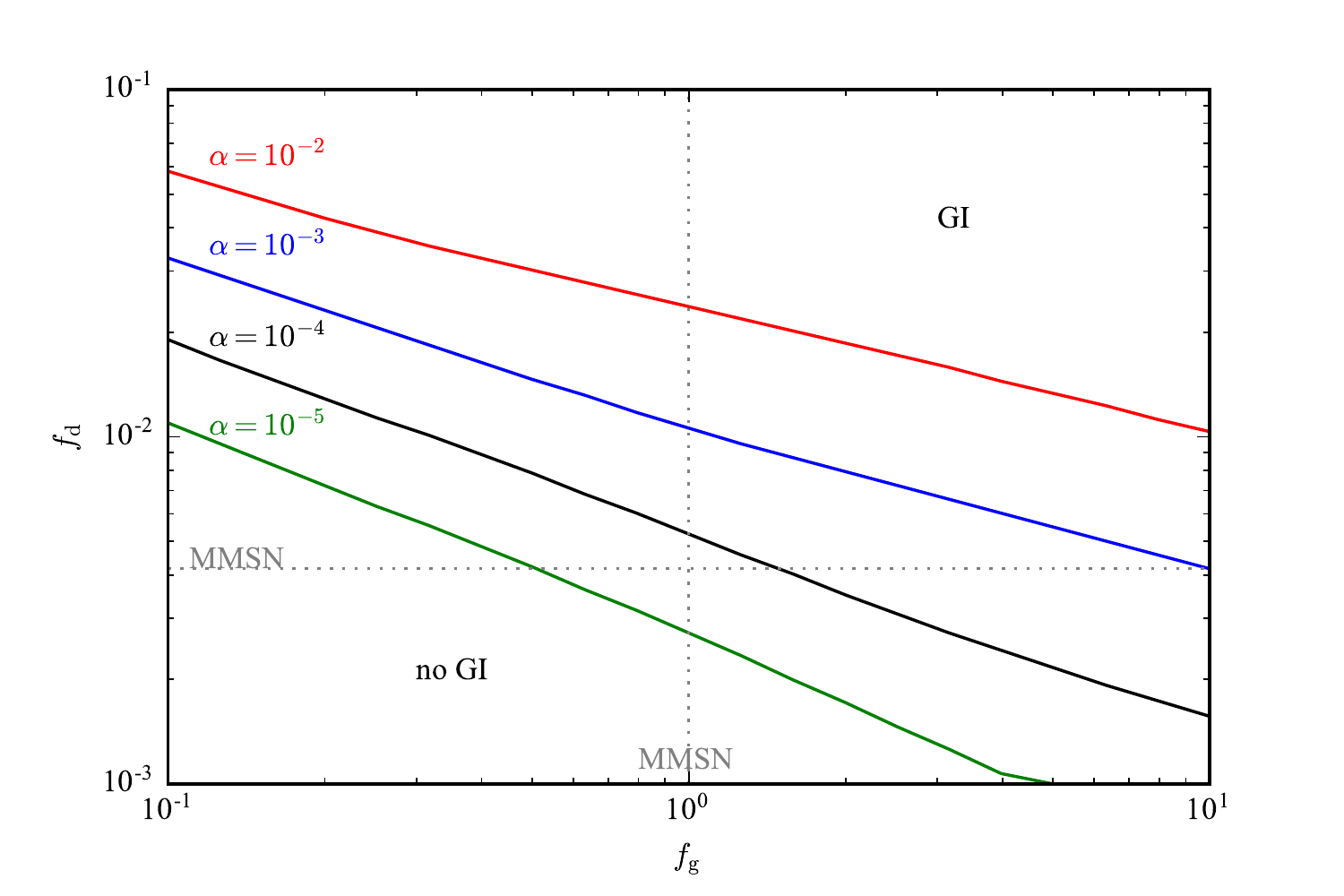}
\caption{Boundaries between the GI and no GI regions in the $f_\mathrm{g}$-$f_\mathrm{d}$ plane.
The orbital radius is 1 au and the dust monomer radius is $r_0=2.5$ nm.
The red, blue, black, and green lines show boundaries for $\alpha=10^{-2}$, $10^{-3}$, $10^{-4}$, and $10^{-5}$, respectively.
The dotted lines show the MMSN model: $f_\mathrm{g}=1$ and $f_\mathrm{d}=0.0042$.}
\label{fig:dependalpha}
\end{figure}

Figure \ref{fig:dependalpha} shows the GI and no GI regions in the $f_\mathrm{g}$-$f_\mathrm{d}$ plane.
The orbital radius and the dust monomer radius are fixed at 1 au and $r_0=2.5$ nm, respectively.
We vary the turbulent strength $\alpha$ and draw boundaries between the two regions.
The GI is found to occur easily in the weak turbulence, massive, and dust-rich disks.
For example, the GI occurs when $\alpha\lesssim10^{-5}$ in the MMSN model at 1 au.

The reason why the GI occurs easily in the massive and/or dust-rich disk is that $Q$ decreases as the dust surface density increases in equation (\ref{eq:Q}).
In the case of the weak turbulent disk, the main increasing mechanism at $Q=2$ is stirring by gas turbulence (Figure \ref{fig:heat}).
When the effect of stirring decreases, the equilibrium random velocity of dust aggregates also decreases, and then, $Q$ decreases.

\subsubsection{Orbital Radius}

\begin{figure}[htbp]
\plotone{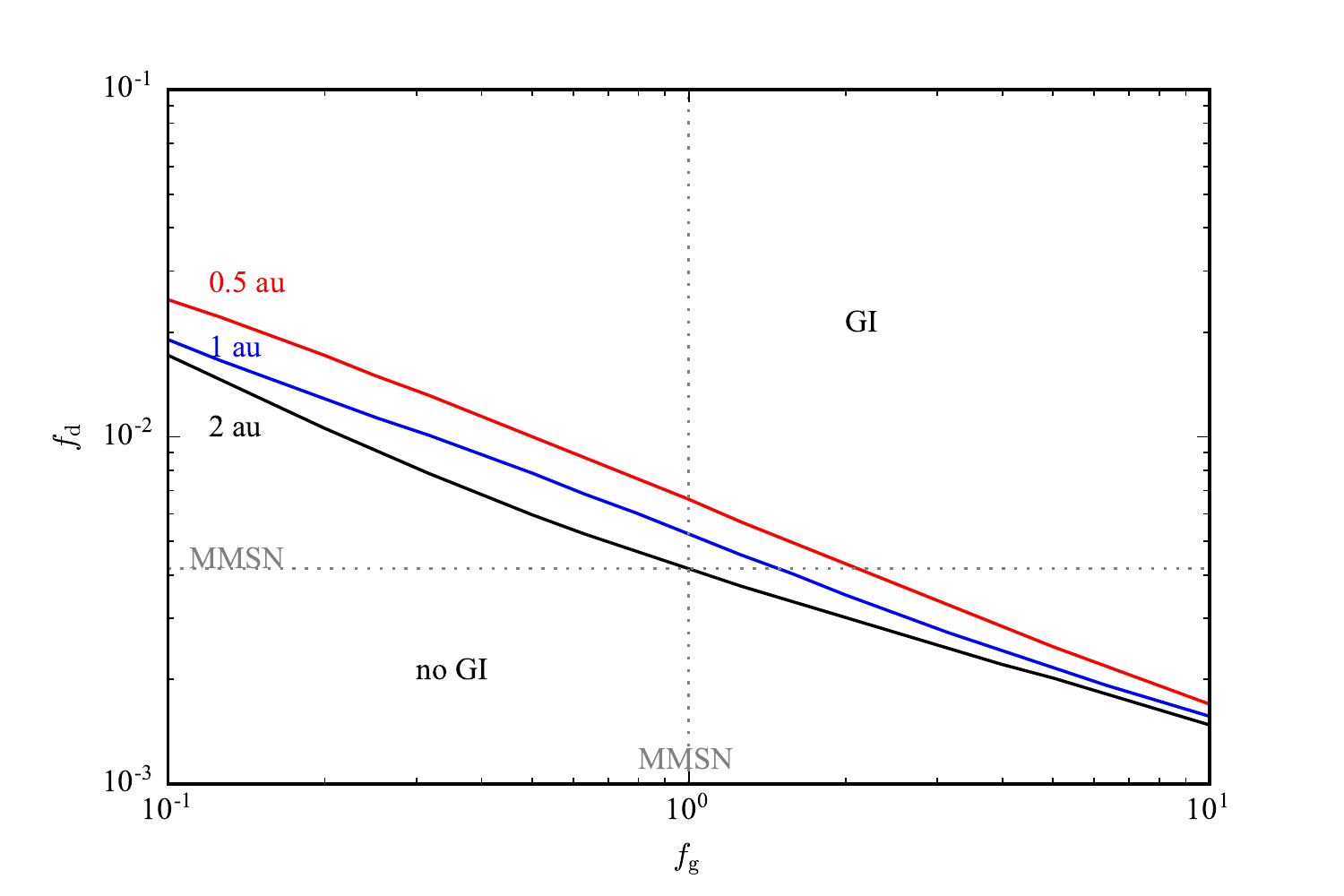}
\caption{Boundaries between the GI and no GI regions in the $f_\mathrm{g}$-$f_\mathrm{d}$ plane.
The turbulent strength is $\alpha=10^{-4}$ and the dust monomer radius is $r_0=2.5$ nm.
The red, blue, and black lines show boundaries for $a=0.5$ au, $a=1$ au, and $a=2$ au, respectively.
The dotted lines show the MMSN model: $f_\mathrm{g}=1$ and $f_\mathrm{d}=0.0042$.}
\label{fig:dependau}
\end{figure}

Next, we vary the orbital radius $a$ and draw boundaries between the GI and no GI regions in Figure \ref{fig:dependau}.
The turbulent strength and the dust monomer radius are fixed at $\alpha=10^{-4}$ and $r_0=2.5$ nm, respectively.
The GI is found to occur more easily at larger orbital radius.
For example, the GI occurs when the orbital radius is larger than 2 au in the MMSN model with $\alpha=10^{-4}$.

The dependence of $Q$ on $a$ is the same as that of the equilibrium random velocity $v_x$ on $a$ because $Q\propto v_x\Omega_\mathrm{K}\Sigma_\mathrm{d}^{-1}\propto v_xa^{-3/2}a^{3/2}\propto v_x$ (equation (\ref{eq:Q})).
At the disk's outer region, the turbulent velocity decreases because of the low temperature, which leads to the weak effect of stirring by gas turbulence.
As a result, the equilibrium random velocity and $Q$ decreases because stirring dominates other increasing mechanisms at $Q=2$.

\subsubsection{Dust Parameters}\label{subsec:resultdust}

\begin{figure}[htbp]
\plotone{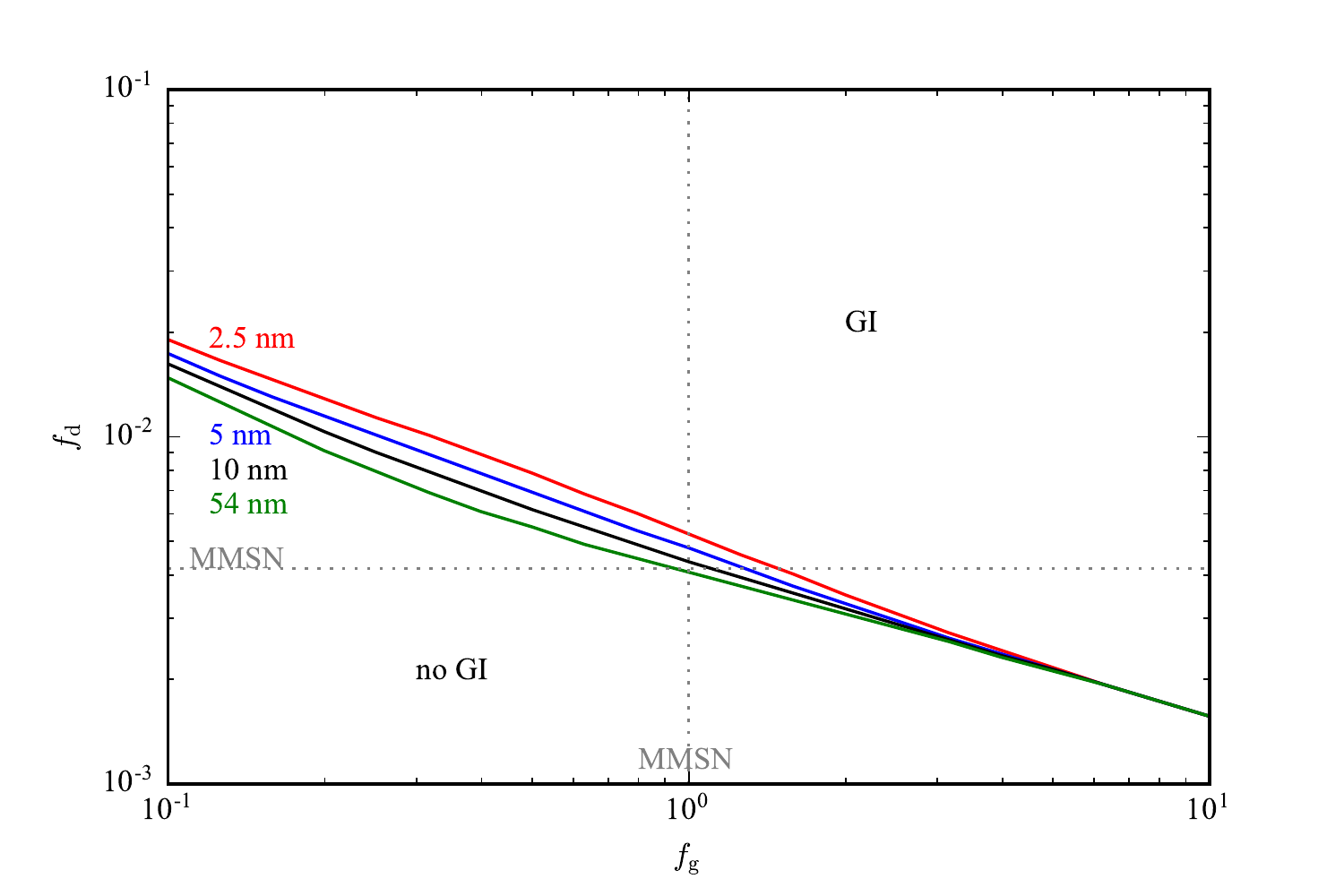}
\caption{Boundaries between the GI and no GI regions in the $f_\mathrm{g}$-$f_\mathrm{d}$ plane.
The turbulent strength is $\alpha=10^{-4}$ and the orbital radius is 1 au.
The red, blue, black, and green lines show boundaries for $r_0=2.5$ nm, $r_0=5$ nm, $r_0=10$ nm, and $r_0=54$ nm, respectively.
The dotted lines show the MMSN model: $f_\mathrm{g}=1$ and $f_\mathrm{d}=0.0042$.}
\label{fig:monomersize}
\end{figure}

Finally, we investigate the dependence of the GI region on the dust monomer radius $r_0$ in Figure \ref{fig:monomersize}.
The turbulent strength and the orbital radius are fixed at $\alpha=10^{-4}$ and $a=1$ au, respectively.
We find that the GI occurs more easily when the dust monomer radius is larger.
In the MMSN model with $\alpha=10^{-4}$ at 1 au, for example, the GI does not occur when the dust monomer radius is less than 10 nm.
However, it becomes more difficult for dust aggregates to stick and grow as the monomer radius becomes larger.
The maximum monomer radius is determined by comparing the maximum collision velocity and the critical velocity of catastrophic disruption $v_\mathrm{frag,cr}$.
Assuming that the maximum collision velocity is the same as the turbulent velocity $v_\mathrm{t}=\sqrt{\alpha}c_\mathrm{s}$, the condition for dust growth is $\sqrt{\alpha}c_\mathrm{s}<v_\mathrm{frag,cr}$.
Thus the monomer radius must be
\begin{equation}
r_0<54\left(\frac{\alpha}{10^{-4}}\right)^{-3/5}\left(\frac{a}{1\mathrm{\ au}}\right)^{3/10}\mathrm{\ nm}.
\end{equation}

\begin{figure}[htbp]
\plotone{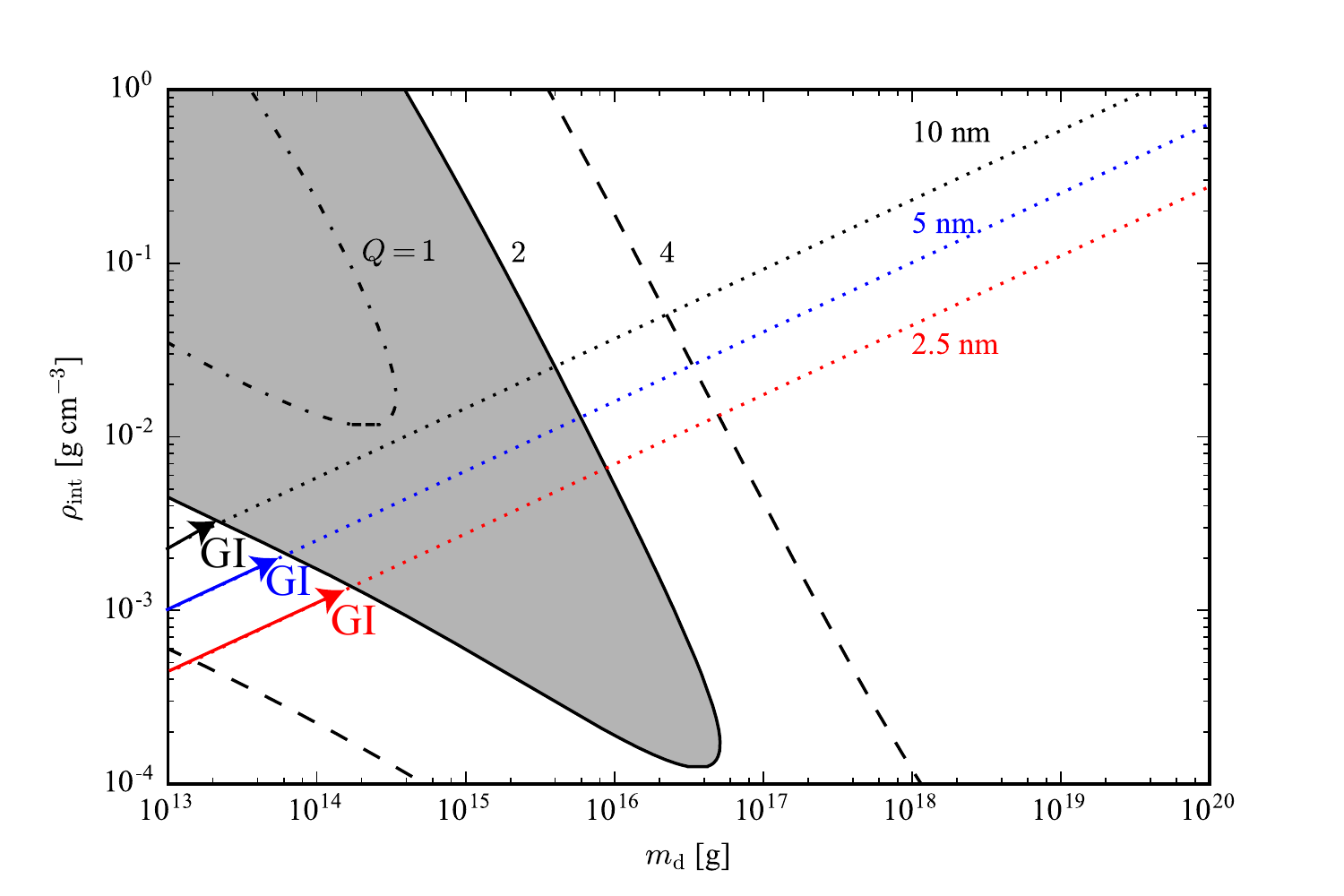}
\caption{Toomre's $Q$ in the $m_\mathrm{d}$-$\rho_\mathrm{int}$ plane at 1 au of the MMSN weak turbulence disk model.
The dash-dotted, solid, and dash contours correspond to $Q=1$, 2, and 4, respectively.
The red, blue, and black dotted lines show the mass and internal density relation under gravitational compression of dust aggregates with $r_0=2.5$, 5, and 10 nm, respectively.}
\label{fig:Qandmonomer}
\end{figure}

Figure \ref{fig:Qandmonomer} shows how the mass and internal density relation of dust aggregates changes in the $m_\mathrm{d}$-$\rho_\mathrm{int}$ plane.
Only the MMSN weak turbulence disk model is shown here because this relation is independent of disk models.
The mean internal density of the dust aggregates increases as their monomer radius increases.
This leads to the weak effect of stirring by gas turbulence.
Finally, the equilibrium random velocity and $Q$ decreases because the main increasing mechanism at $Q=2$ is stirring.

\section{Conclusions}\label{sec:sum}

We have investigated the gravitational instability (GI) of a dust layer composed of porous dust aggregates of $\sim2.5$--$10$-nm-sized silicate monomers.
To evaluate the disk stability, we calculated Toomre's stability parameter $Q$ from the equilibrium random velocity of dust aggregates.
We calculated the equilibrium random velocity considering five processes: gravitational scattering between dust aggregates, collisions between them, drag by mean flow of gas, stirring by gas turbulence, and gravitational scattering by gas density fluctuation due to turbulence.

We derived the GI condition as a function of five disk and dust parameters: disk mass, dust-to-gas ratio, turbulent strength, orbital radius, and dust monomer radius.
In the case of the minimum mass solar nebula model at 1 au, for example, the dust layer becomes gravitationally unstable when the turbulent strength $\alpha\lesssim10^{-5}$ and the monomer radius $r_0=2.5$ nm.
If the dust-to-gas ratio is increased twice, the GI occurs for $\alpha\lesssim10^{-4}$.
We found that the GI occurs more easily in the more massive and more dust-rich disks with weaker turbulence at outer regions.
The larger monomer radius is preferable to the GI.

In this paper, we only investigated the condition of $Q<2$, which leads to the growth of self-gravity wakes in a dust layer.
However, it is unknown how such wakes fragment to form planetesimals.
For the fragmentation of a gas disk, the cooling timescale should be comparable to or shorter than the orbital timescale \citep{Gammie2001}.
\cite{Michikoshi2017} pointed out the existence of a similar condition for the wake fragmentation.
However to clarify this condition is out of the present scope and will be a future work.
If planetesimals are formed in the self-gravity wakes, their typical mass $m_\mathrm{p}$ can be estimated by
\begin{equation}
m_\mathrm{p}\simeq\lambda_\mathrm{cr}^2\Sigma_\mathrm{d}\simeq1.6\times10^{18}f_\mathrm{g}^3\left(\frac{f_\mathrm{d}}{0.0042}\right)^3\left(\frac{a}{1 \mathrm{\ au}}\right)^{3/2}\mathrm{\ g},
\end{equation}
where $\lambda_\mathrm{cr}=4\pi^2G\Sigma_\mathrm{d}/\Omega_\mathrm{K}^2$ is the critical wavelength of the GI.

In addition, the rolling energy $E_\mathrm{roll}$ of silicate monomers has a large uncertainty because the critical displacement $\xi$ is different between theoretical and experimental values.
Note that the dependence on $\xi$ is weaker than that on the monomer radius $r_0$ as suggested by equation (\ref{eq:evoltrack}).

Moreover, the assumption of single dust aggregate radius $r_\mathrm{d}$ may not be appropriate even if the runaway growth does not occur.
In reality, the dust aggregates have the size distribution due to collisional fragmentation.
The effects of the dust size distribution on the GI have to be investigated.

\acknowledgments{
We thank Akimasa Kataoka and Tetsuo Taki for fruitful discussions.
We appreciate the careful reading and valuable comments by the anonymous referee and the editor, Judy Pipher.
}

\bibliography{paper}

\begin{thebibliography}{}
\expandafter\ifx\csname natexlab\endcsname\relax\def\natexlab#1{#1}\fi

\bibitem[{{Adachi} {et~al.}(1976){Adachi}, {Hayashi}, \&
  {Nakazawa}}]{Adachi1976}
{Adachi}, I., {Hayashi}, C., \& {Nakazawa}, K. 1976, Progress of Theoretical
  Physics, 56, 1756

\bibitem[{{Andrews} {et~al.}(2010){Andrews}, {Wilner}, {Hughes}, {Qi}, \&
  {Dullemond}}]{Andrews2010}
{Andrews}, S.~M., {Wilner}, D.~J., {Hughes}, A.~M., {Qi}, C., \& {Dullemond},
  C.~P. 2010, \apj, 723, 1241

\bibitem[{{Arakawa} \& {Nakamoto}(2016)}]{Arakawa2016}
{Arakawa}, S., \& {Nakamoto}, T. 2016, \apjl, 832, L19

\bibitem[{{Blum} \& {M{\"u}nch}(1993)}]{Blum1993}
{Blum}, J., \& {M{\"u}nch}, M. 1993, \icarus, 106, 151

\bibitem[{{Blum} \& {Wurm}(2000)}]{Blum2000}
{Blum}, J., \& {Wurm}, G. 2000, \icarus, 143, 138

\bibitem[{{Brown} \& {Lawler}(2003)}]{Brown2003}
{Brown}, P.~P., \& {Lawler}, D.~F. 2003, Journal of Environmental Engineering,
  129, 222

\bibitem[{{Dominik} \& {Tielens}(1997)}]{Dominik1997}
{Dominik}, C., \& {Tielens}, A.~G.~G.~M. 1997, \apj, 480, 647

\bibitem[{{Gammie}(2001)}]{Gammie2001}
{Gammie}, C.~F. 2001, \apj, 553, 174

\bibitem[{{Goldreich} \& {Ward}(1973)}]{Goldreich1973}
{Goldreich}, P., \& {Ward}, W.~R. 1973, \apj, 183, 1051

\bibitem[{{Hayashi}(1981)}]{Hayashi1981}
{Hayashi}, C. 1981, Progress of Theoretical Physics Supplement, 70, 35

\bibitem[{Heim {et~al.}(1999)Heim, Blum, Preuss, \& Butt}]{Heim1999}
Heim, L.-O., Blum, J., Preuss, M., \& Butt, H.-J. 1999, Phys. Rev. Lett., 83,
  3328

\bibitem[{{Ida}(1990)}]{Ida1990}
{Ida}, S. 1990, \icarus, 88, 129

\bibitem[{{Inaba} {et~al.}(2001){Inaba}, {Tanaka}, {Nakazawa}, {Wetherill}, \&
  {Kokubo}}]{Inaba2001}
{Inaba}, S., {Tanaka}, H., {Nakazawa}, K., {Wetherill}, G.~W., \& {Kokubo}, E.
  2001, \icarus, 149, 235

\bibitem[{{Kataoka} {et~al.}(2013{\natexlab{a}}){Kataoka}, {Tanaka}, {Okuzumi},
  \& {Wada}}]{Kataoka2013L}
{Kataoka}, A., {Tanaka}, H., {Okuzumi}, S., \& {Wada}, K. 2013{\natexlab{a}},
  \aap, 557, L4

\bibitem[{{Kataoka} {et~al.}(2013{\natexlab{b}}){Kataoka}, {Tanaka}, {Okuzumi},
  \& {Wada}}]{Kataoka2013}
---. 2013{\natexlab{b}}, \aap, 554, A4

\bibitem[{{Keller} \& {Messenger}(2011)}]{Keller2011}
{Keller}, L.~P., \& {Messenger}, S. 2011, \gca, 75, 5336

\bibitem[{{Kokubo} \& {Ida}(1996)}]{Kokubo1996}
{Kokubo}, E., \& {Ida}, S. 1996, \icarus, 123, 180

\bibitem[{{Michikoshi} {et~al.}(2007){Michikoshi}, {Inutsuka}, {Kokubo}, \&
  {Furuya}}]{Michikoshi2007}
{Michikoshi}, S., {Inutsuka}, S.-i., {Kokubo}, E., \& {Furuya}, I. 2007, \apj,
  657, 521

\bibitem[{{Michikoshi} \& {Kokubo}(2016)}]{Michikoshi2016GI}
{Michikoshi}, S., \& {Kokubo}, E. 2016, \apjl, 825, L28

\bibitem[{{Michikoshi} \& {Kokubo}(2017)}]{Michikoshi2017}
---. 2017, \apj, 842, 61

\bibitem[{{Okuzumi} \& {Ormel}(2013)}]{Okuzumi2013}
{Okuzumi}, S., \& {Ormel}, C.~W. 2013, \apj, 771, 43

\bibitem[{{Okuzumi} {et~al.}(2012){Okuzumi}, {Tanaka}, {Kobayashi}, \&
  {Wada}}]{Okuzumi2012}
{Okuzumi}, S., {Tanaka}, H., {Kobayashi}, H., \& {Wada}, K. 2012, \apj, 752,
  106

\bibitem[{{Sekiya}(1983)}]{Sekiya1983}
{Sekiya}, M. 1983, Progress of Theoretical Physics, 69, 1116

\bibitem[{{Sekiya}(1998)}]{Sekiya1998}
---. 1998, \icarus, 133, 298

\bibitem[{{Shakura} \& {Sunyaev}(1973)}]{Shakura1973}
{Shakura}, N.~I., \& {Sunyaev}, R.~A. 1973, \aap, 24, 337

\bibitem[{{Stewart} \& {Ida}(2000)}]{Stewart2000}
{Stewart}, G.~R., \& {Ida}, S. 2000, \icarus, 143, 28

\bibitem[{{Suyama} {et~al.}(2008){Suyama}, {Wada}, \& {Tanaka}}]{Suyama2008}
{Suyama}, T., {Wada}, K., \& {Tanaka}, H. 2008, \apj, 684, 1310

\bibitem[{{Toomre}(1964)}]{Toomre1964}
{Toomre}, A. 1964, \apj, 139, 1217

\bibitem[{{Toomre}(1981)}]{Toomre1981}
{Toomre}, A. 1981, in Structure and Evolution of Normal Galaxies, ed. S.~M.
  {Fall} \& D.~{Lynden-Bell}, 111--136

\bibitem[{{Toriumi}(1989)}]{Toriumi1989}
{Toriumi}, M. 1989, Earth and Planetary Science Letters, 92, 265

\bibitem[{{Wada} {et~al.}(2007){Wada}, {Tanaka}, {Suyama}, {Kimura}, \&
  {Yamamoto}}]{Wada2007}
{Wada}, K., {Tanaka}, H., {Suyama}, T., {Kimura}, H., \& {Yamamoto}, T. 2007,
  \apj, 661, 320

\bibitem[{{Wada} {et~al.}(2008){Wada}, {Tanaka}, {Suyama}, {Kimura}, \&
  {Yamamoto}}]{Wada2008}
---. 2008, \apj, 677, 1296

\bibitem[{{Wada} {et~al.}(2009){Wada}, {Tanaka}, {Suyama}, {Kimura}, \&
  {Yamamoto}}]{Wada2009}
---. 2009, \apj, 702, 1490

\bibitem[{{Youdin}(2011)}]{Youdin2011}
{Youdin}, A.~N. 2011, \apj, 731, 99

\bibitem[{{Youdin} \& {Lithwick}(2007)}]{Youdin2007}
{Youdin}, A.~N., \& {Lithwick}, Y. 2007, \icarus, 192, 588

\end{thebibliography}

\end{document}